multipoles%12/11/2001
\documentstyle[12pt,epsfig,graphics]{article}
\newcommand{\be}{\begin{equation}}
\newcommand{\ee}{\end{equation}}
\newcommand{\ba}{\begin{eqnarray}}
\newcommand{\ea}{\end{eqnarray}}
\newcommand{\lb}{\label}
\newcommand{\bb}{\bibitem}
\newcommand{\half}{\frac{1}{2}}

\newcommand{\nn}{\nonumber}

\begin{document}

\begin{center}
{\LARGE  Relativistic Multipoles and the  Advance of the\\

\vspace*{0.2cm} 
Perihelia}

\vspace{1cm} 

 {\large \em Bruno Boisseau\footnote{e-mail:
 boisseau@celfi.phys.univ-tours.fr } { \em and} 
  Patricio S. Letelier\footnote{e-mail: 
letelier@ime.unicamp.br}$^{,}$\footnote{Permanent
 address: Departamento
 de Matem\'atica Aplicada-IMECC,
Universidade Estadual de Campinas,
13083-970 Campinas. S.P., Brazil   }  }
\vspace{1ex} 
 
 Laboratoire de Math\'ematiques et Physique Th\'eorique\\
 CNRS/UMR 6083, Universit\'e Fran\c{c}ois Rabelais\\
 Facult\'e des Sciences et Techniques\\             
 Parc de Grandmont 37200 TOURS, France
\end{center}  

\vspace{6ex} 

\centerline{ \small\bf Abstract}
   
\baselineskip 0.7cm  
 In order to shed some light in the meaning of the relativistic
 multipolar expansions we consider different static solutions
 of the axially symmetric vacuum 
Einstein equations that  in  the non relativistic limit
 have  same Newtonian moments. 
  The motion of  test particles orbiting around different deformed attraction
 centers with the same Newtonian limit  is  studied paying
 special attention to the advance of
 the perihelion.  We 
find  discrepancies  in the fourth order of the
dimensionless  parameter (mass of the attraction center)/(semilatus
rectum). An  evolution  equation for the difference
 of the radial
coordinate due to the use of  different general relativistic
 multipole expansions is
presented.


\newpage

\section{Introduction.}

The adequate description of the gravitation field of an astrophysical
object has been an important subject in both relativistic and Newtonian 
gravity since their origins. The  particular case of the gravity associated 
to  axially symmetric bodies  has played a central role in this discussion. 
Recently, Merrit \cite{merr} found,  from detailed  modeling of triaxial 
galaxies, that most of the galaxies must be nearly axisymmetric, either
 prolate or oblate. In Newtonian theory the gravitational potential of 
axially symmetric bodies  can be always represented by its usual expansion
 in terms of Legendre polynomials (zonal harmonics). In general relativity
 we have that the  solution of the vacuum Einstein equations associated
 to  static  axially symmetric bodies has a simple form with only
 two metric functions \cite{W}.

 As a consequence of   the self interaction of the
 general relativistic gravitational field we  have a  rather  elaborate 
 description of   multipolar fields,  the  covariant multipole
 moments \cite{kip}.
For the special case of axially symmetric spacetimes
 we have   geometric properties that can also  be used 
to define a class of multipole moments \cite{bob,han}.
 It is highly non  trivial   to prove the equivalence between these two
definitions  \cite{gursel}.
Also,  due to the nonlinearity of the Einstein equations we do not have a simple 
 expansion  of the metric
in terms of relativistic multipoles like the expansion of the potential
 in Newtonian gravity. Moreover, 
it is not know 
  how to construct a solution to the vacuum Einstein equation with
 assigned relativistic multipoles. Also the physical consequences of
 having different relativistic moments are rather obscure.
Furthermore,  the problem to model a static  axially symmetric center
 of attraction in
 general relativity with some multipolar perturbations   that have a prescribed
 Newtonian limit is not a well define problem, as a matter of fact we have
 an infinite number of solutions  of the vacuum Einstein equations with the
 same Newtonian limit.  
This problem arises in a natural form when one studies the
  orbits in
deformed centers of attraction, see for instance \cite{guelet}.

 In this paper,
 in order to shed some light in the physical meaning of the relativistic
 multipolar expansions,  we consider different static solutions
 of the axially symmetric vacuum 
Einstein equations that  in  the non relativistic limit
 have  same Newtonian moments. The particular solutions 
considered represent  a large 
class of rather natural solutions that  are generalizations of
 some well know ones.

Along this article we shall use the terminology ``distorted black hole'' 
introduced in \cite{z} to indicate an  attraction center with multipolar
moments, examples are: a) A true black hole (or a dense object)
 surrounded by  a distribution of matter like a ring
or a small disk  formed by counterrotating matter, i.e., built
by  approximately the same number of  particles
 moving  clockwise as counterclockwise.  Even though, this
 interpretation can be 
seen as a device to have a static stable configuration
there are observational evidence of disks 
made of streams of  rotating and counterrotating matter \cite{counter}.
b) An axially symmetric static dense object with either polar deformations
or polar jets.
In our analysis the horizon will play no role, therefore
 our results will be  valid
for any  nearly static axially symmetric  attraction center.

 Since the advance 
 of the perihelion is one of the most  significant features of the general 
relativistic celestial mechanics will be considered in some detail.
 We 
find  discrepancies  in the fourth order of the
dimensionless  parameter (mass of the attraction center)/(semilatus
rectum). We present an  evolution  equation for the difference
 of the radial
coordinate due to the use of  different general relativistic
 multipole expansions.

In Sec. 2 we present a summary of the main expressions associated
 to the Weyl solutions. In Sec. 3 we study the Newtonian limit of 
four different  solutions to the Einstein equations.
 Also we study the relativistic
multipoles moments for  each solution.
These solutions are the Erez-Rosen-Quevedo (ERQ)
 solution \cite{que}, a solution characterized by the usual Newtonian
 multipoles and generalizations of both. We present two graphics
that show the difference between the solutions.
In the next section, Sec. 4, we study series expansions of the
 solutions restricted to the equatorial plane. Also several
 differences of the metric
 potentials are considered. In Sec. 5 an evolution equation 
for the difference of the radial coordinate due to the use of  different 
multipole expansions is presented. We solve this equation for circular orbits
in the equatorial plane.
In Sec. 6, the perihelion shift is computed to the order in which the first 
discrepancies appears due to the use of the different multipolar expansions.
As a particular case we have the perihelion shift in a Schwarzschild
 spacetime up to a fourth order. We also solve the difference equation of
 the preceding section for generic orbits on the equatorial plane.
Finally, in Sec. 7, we present a short  discussion of our main results.

\section{Static axially symmetric solutions of Einstein vacuum equations.}

 The external gravitational field of produced by an axially symmetric
body can be well  described by the Weyl metric  \cite{W},
\be
ds^2=e^{2\psi}dt^2 -e^{-2\psi}[r^2 d\varphi^2+e^{2\gamma}(dr^2+dz^2)],
 \lb{weylm}
\ee
where the functions $\psi$ and $\gamma$ depend only on $r$ and $z$; the
 ranges of the coordinates  $(r,\varphi,z)$ are the usual for cylindrical
 coordinates and $-\infty<t<+\infty$. The vacuum Einstein equations ($
R_{\mu \nu}=0 $) reduces to the usual Laplace equation in
 cylindrical coordinates,
\be
\psi_{,rr}+\psi_{,r}/r+\psi_{,zz}=0, \lb{lapw} 
\ee
and the quadrature,
\be 
d\gamma[\psi]= r[(\psi_{,r}^2-\psi_{,z}^2)dr+2\psi_{,r}\psi_{,z}dz].
 \lb{gawc}
\ee
When  $\psi$ satisfies the Laplace equation this differential is exact. 

   The Schwarzschild solution in Weyl coordinates takes the form
 \ba
\psi_S&=&\frac{1}{2}\ln\frac{R_++R_- -2m}{R_++R_-+2m},  \lb{psisw}\\
\gamma_S&=&\half\ln\frac{(R_++R_-)^2-4m^2}{R_+R_-}, \lb{gasw}
\ea
where
\be
R_\pm=\sqrt{r^2+(z\pm m)^2}. \lb{Rpm}
\ee
The function $\psi_S$ is just the Newtonian potential of a  bar  of
 length $2m$ and density $1/2$. The fact that in Weyl coordinates 
 the Schwarzschild metric does not look
spherically symmetric and that the horizon is squeezed into a line of 
length $2m$ has prevented the wide use of these coordinates.
 Since the superposition of static axially symmetric perturbation 
can be performed in a simple way and exact results can be obtained, 
we believe that is worth  to use these coordinates. More suitable
 coordinates are the spheroidal coordinates,
\be 
x=(R_+ + R_-)/(2m),\;\; y=(R_+ - R_-)/(2m). \lb{pro}
\ee
Note that the coordinates $(x,y)$ are dimensionless and $1\leq x<\infty$
and $ -1\leq y \leq 1 $. The coordinate $x$ is essentially a radial coordinate and $y$ the cosine of an angle. The inverse of (\ref{pro}) are,
\be 
r=m\sqrt{(x^2-1)(1-y^2)},\;\; z=m xy. \lb{cylpro}
\ee
In this case the Weyl metric takes the form,
\be
ds^2=e^{2\psi}dt^2-m^2 e^{-2\psi}[e^{2\gamma}(x^2-y^2)(\frac{dx^2}{x^2-1}
+\frac{dy^2}{1-y^2})+(x^2-1)(1-y^2)d\varphi^2], \lb{mxy}
\ee
and the  Einstein equations,
\ba
&&[(1-x^2)\psi_{,x}]_{,x}+[(y^2-1)\psi_{,y}]_{,y}=0, \lb{lapxy} \\
&&\gamma_{,x}=\frac{1-y^2}{x^2-y^2}[x(x^2-1)\psi_{,x}^2-x(1-y^2)\psi_{,y}^2
 -2y(x^2-1)\psi_{,x}\psi_{,y}], \lb{gax}\\
&&\gamma_{,y}=\frac{x^2-1}{x^2-y^2}[y(x^2-1)\psi_{,x}^2-y(1-y^2)\psi_{,y}^2
 +2y(1-y^2)\psi_{,x}\psi_{,y}]. \lb{gay}
\ea
 In this coordinates the Schwarzschild solutions takes the
 particularly simple
form
\ba
\psi_S &=&-Q_0(x),   \lb{psxy}  \\
&=&\frac{1}{2}\ln\frac{x-1}{x+1}, \nonumber\\
\gamma_S&=&\half \ln\frac{x^2-1}{x^2-y^2}\lb{gasxy}.
\ea
From  the coordinate transformations, $x=R_s/m-1,\;\; y=
\cos\vartheta$, and 
 (\ref{mxy}) with (\ref{psxy}) and (\ref{gasxy}) we recover the 
Schwarzschild metric in its usual  form,
\be
ds^2=(1-2m/R_s)dt^2-(1-2m/R_s)^{-1}dR_s^2- R_s^2(d\vartheta^2+
\sin\vartheta^2 d\varphi^2). \lb{Schw}
\ee

We  shall consider solutions of the form,
\be
\psi=\psi_S +\hat{\psi}, \lb{supp}
\ee
where $\hat{\psi}(x,y)$ represents the superposition of external multipolar
 fields  solutions of Laplace equation with  no monopolar term.
The function $\gamma$ in this case can be written as
\be
\gamma[\psi]=\gamma_S+\hat\gamma,  \lb{supg}
\ee
with
\ba
&&\hat\gamma=2\gamma[\psi_S,\hat\psi]+\gamma[\hat\psi] \lb{gc}, \\
&& d\gamma[\psi_S,\hat\psi]=r[(\psi_{S,r}\hat{\gamma}_{,r}-
\psi_{S,z}\hat{\gamma}_{,z})dr+(\psi_{S,r}\hat{\gamma}_{,z}+
\psi_{S,z}\hat{\gamma}_{,r})dz].
 \lb{gasp} 
\ea
In  this case the Weyl metric takes the form,
\be
ds^2=(1-\frac{2m}{R_S})e^{2\hat\psi}dt^2-
\frac{e^{2(\hat\gamma-\hat\psi)}}
{1-\frac{2m}{R_S}}
dR_S^2 -R_S^2 e^{2(\hat{\gamma}-\hat\psi)}d\vartheta^2 -
R_S^2 e^{-2\hat\psi}\sin^2\vartheta d\varphi^2, \lb{sp}
\ee
that has been  interpreted as a ``black hole''
 with multipolar deformations \cite{z}.  

Note that this metric for a given
 function $\hat \psi$ -- solution of the usual axially symmetric Laplace equation -- and
its associated potential $\hat \gamma$ given by Eq.  (\ref{gc}) is an exact solution of
the Einstein vacuum field equations. We shall consider several different classes of 
approximations. In principle, since we start with an exact framework we will be able
to  control the validity of the approximations. The relation of this Weyl approach
 in the case linear perturbations  with the  Regge-Wheeler \cite{RW} formalism
 --  {\em general linear}  perturbations of a black hole -- 
 can be found in \cite{vlapj}. In the present paper, in some cases, we will 
go beyond the linear case.

\section{Multipoles and multipolar fields.}

In order to gain some understanding of the solutions of the vacuum 
Einstein equations with different relativistic moments,  
in this section,  we  study different Weyl solutions that
  represent  static 
attraction centers that have the same Newtonian limit.

Following Ehlers \cite{ehlers},
 for a given Weyl solution we  define its Newtonian limit as 
 $\lim_{\lambda=0}\psi(\lambda,r,z)/\lambda$,
  where $\lambda \equiv c^{-2}$ and $c$ the
light velocity, e.g.  for the Schwarzschild solution we 
 put  $m=\lambda G M$ in
(\ref{psisw}) and  with the aid of  l'H\^opital rule we get
\be
\lim_{\lambda=0}\psi_S(\lambda,r,z)/\lambda=-GM/R. \lb{N0}
\ee  
with
\be
R=\sqrt{r^2+z^2}. \lb{R}
\ee

There are an infinite number of different Weyl solutions that represent
  fields of
multipoles with the same  Newtonian multipolar  limit. We shall 
restrict our study
to two known solutions that have been  used recently
 in some applications
 and other two that are generalizations of the formers. They are:

a) The  usual Newtonian multipolar fields,
\be
\hat{\psi}^N=\sum_{k=1}^\infty q_k \frac{P_k(z/R)}{R^{k+1}}, \lb{psing}
\ee
we have the same $\hat{\psi}^N$ as its Newtonian limit. Note that
 $q_k$ scales with $\lambda$ as  $\lambda q_k$.

b) Separating variables in (\ref{lapxy}) one finds the ERQ
solution \cite{ER,que},
\be
\hat{\psi}^{ERQ}=\sum_{k=1}^\infty\frac{(2k+1)!}
{2^k (k!)^2 m^{k+1}} q_k Q_k(x)P_k(y), \lb{psierqg}
\ee
where $Q_k(x)$  and $P_k(y)$ are the Legendre functions of the
 second kind and the Legendre polynomials, respectively. We have chosen
 the coefficients in the series to have
\be 
\lim_{\lambda=0}\psi^{ERQ}_n(\lambda,r,z)/\lambda=\psi^N_n, \lb{nlnerq}
\ee
where   $\psi^{ERQ}_n$ and $\psi^N_n$ are the $n$-term in the series 
(\ref{psierqg}) and (\ref{psing}), respectively. To find the limit is useful the Letelier
identity \cite{letid},
\be
Q_n(x)P_n(y)=\frac{1}{2}\int_{-m}^{m}\frac{P_n(\alpha/m)}
{\sqrt{r^2+(z-\alpha)^2}}d\alpha . \lb{letid}
\ee 

c) The solutions of Laplace equation:
\ba
&&\hat{\psi}^I=\sum_{k=1}^\infty q_k[\alpha
 \frac{ P_n((z+am)/R_{+a})}{R_{+a}^{k+1}} +
(1-\alpha)\frac{P_n((z-am)/R_{-a})}{R_{-a}^{k+1}}] , \lb{psIg}\\
&&\hat{\psi}^{II}=\sum_{k=1}^\infty\frac{(2k+1)!}
{2^k (k!)^2 (ma)^{k+1}} q_k Q_k(x_a)P_k(y_a), \lb{psIIg}
\ea
where
\ba 
&&x_a=(R_{+a} + R_{-a})/(2ma),\;\; y_a=(R_{+a} - R_{-a})/(2ma), \lb{proa}\\
&& R_{+a}=\sqrt{r^2+(z+am)^2},\;\; R_{-a}=\sqrt{r^2+(z-am)^2}, \lb{Rmma}
\ea
$\alpha$ and $a$ are arbitrary constants that we shall take  as
positive and less than one, we shall comeback to this point later. 
$\hat{\psi}^I$ is obviously a solution of 
the Laplace equation.  To prove that $\hat{\psi}^{II}$ is also a solution of
this last equations one can use the identity (\ref{letid}).  $\hat{\psi}^I$ 
and $\hat{\psi}^{II}$  are nontrivial deformations of the Newtonian multipoles
and the ERQ solution, respectively. Note that in the superposition (\ref{supp})
the Schwarzschild term $\psi_S$  is not changed. 
Letting $a=0$ in 
$\psi^I$ we get $\psi^N$, also putting $a=1$ in $\psi^{II}$ we recover 
$\psi^{ERQ}$. The global properties of
these new  solutions will be presented elsewhere. 

The first two solutions are the  axially symmetric multipolar
 expansion for the Newtonian
 Laplace equation in spherical coordinates (usual expansion) and in spheroidal
 prolate coordinates, respectively. The  second 
one has the property of having the monopolar  term  proportional to the Weyl
 potential that
gives rise to the Schwarzschild solution [cf. Eq. (\ref{psxy})]. The 
 Schwarzschild solution 
with multipolar deformations  represented by either expansions have been
 study by several
authors (see for instance, \cite{que,manko}),  specially in the important case of 
quadrupolar deformations \cite{z, ER}. For  recent applications 
 see \cite{guelet,herrera}.
The last two multipolar expansions, $\hat{\psi}^I$ and  $\hat{\psi}^{II},$
are new ones and are closed related to the previous ones. They have some new parameters that 
do not appear in the Newtonian limit. They will be  considered  in order to
  explore the possibility to have relativistic multipoles not completely 
determined by its Newtonian ones. We shall  comeback to this point later.

The relativistic multipoles $m_k$  for a Weyl solution
can be computed from the function $\psi$ evaluated on
 the axis of symmetry, in the following way \cite{que,hoen}: First we define the function
\be
\xi(x,y)=\frac{1-e^{2\psi(x,y)}}{1+e^{2\psi(x,y)}}\label{xi}
\ee
and find its value on the axis of symmetry $r=0$, i.e., 
$y=1$; so $x=z/m$. Then we calculate  the quantities
\be
\bar m_k=\frac{1}{(k+1)!}\frac{d^{k+1}\xi(\tilde{z},1)} 
{d \tilde{z}^{k+1}}|_{\tilde{z}=0} \lb{mk}
\ee 
with $\tilde{z}=1/z$.
The relativistic moments are given by
\be
m_k=\bar{m}_k+d_k \lb{Mk},
\ee
where the first six constants $d_k$ are \cite{hoen}:
\ba
&& d_0=d_1=d_2=d_3=0, \nonumber\\
&& d_4=\bar m_0(\bar m_1^2-\bar m_2 \bar m_0)/4, \nonumber\\
&& d_5=\bar m_0(\bar m_2 \bar m_1 -\bar m_3 \bar m_0)/3+(\bar m_1(\bar m_1^2-
\bar m_2 \bar m_0)/21. \lb{dk}
\ea

The direct computations of the relativistic multipolar moments up to
 the sixteenth pole gives:

a) For the Newtonian multipolar field $\hat{\psi}^N$,
\ba
&& m_0=m, \;\;\ m_1=q_1, \;\; m_2=q_2, \;\; m_3=q_3-m^2 q_1, \lb{rmN}\\
&& m_4=q_4 -(8 m^2 q_2 + 6 m q_1^2)/7.  \nonumber 
\ea

b) For the ERQ multipolar field $\hat{\psi}^{ERQ}$,
\ba
&& m_0=m, \;\;\ m_1=q_1, \;\; m_2=q_2, \;\; m_3=q_3-2m^2 q_1/5, \lb{rmERQ}\\
&& m_4=q_4 -2( m^2 q_2 + 3 m q_1^2)/7.  \nonumber
\ea

c) For the  multipolar field $\hat{\psi}^{I}$,
\ba
&& m_0=m, \;\;\ m_1=q_1, \;\; m_2=q_2 +2(1-2\alpha)a m q_1,  \lb{rmI}\\
&& m_3=q_3+(3a^2-1)m^2q_1+3a(1-2\alpha)mq_2 \nonumber\\
&& m_4=q_4 +4a(7a^2-4)(1-2\alpha)m^3q_1/7\nonumber\\
&& \hspace{1cm}+4(1-2\alpha)amq_3+2(21a^2-4)m^2q_2/7 
-6mq_1^2/7 .\nonumber
\ea

d) For the  multipolar field $\hat{\psi}^{II}$
\ba
&& m_0=m, \;\;\ m_1=q_1, \;\; m_2=q_2, \;\; m_3=q_3-(1-3a^2/5)m^2 q_1,
 \lb{rmII}\\
&& m_4=q_4 -6mq_1^2/7 -2(4-3a^2)m^2q_2/7.  \nonumber
\ea

Since our main goal is to  study the physical effect of
having different relativistic multipole moments is instructive to
compute the differences of these multipoles for the above mentioned,
 solutions. We shall compute the 
quantities,
\be
\Delta m_k= m_k-m^N_k. \lb{dm}
\ee
We find for the ERQ, I and II solutions $\Delta m_0=\Delta m_1=0$, and

\noindent
a) For the ERQ solution:
\be 
\Delta m_2=0,\;\;  \Delta m_3 =3mq_1/5, \;\;  \Delta m_4 =6m^2q_2/7. \lb{dmerq}
\ee
b) For the solution I:
 \ba
&& \Delta m_2 =2a(1-2\alpha)mq_1, \;\;\Delta m_3 =3a^2m^2q_1+3a(1-2\alpha)mq_2, \nonumber\\
&&\Delta m_4 =4a(1-2\alpha)(7a^2-4)m^3q_1/7+ 6a^2m^2q_2+4a(1-2\alpha)mq_3. \lb{dmI}
\ea
c) For the solution II:
\be
 \Delta m_2=0,\;\;  \Delta m_3 =3a^2m^2q_1/5, \;\;  \Delta m_4 =6a^2m^2q_2/7.
  \lb{dmII}
\ee
Along all  the paper we shall compare the different solutions with $\psi^N$, of course this choice is arbitrary.

Since  the quadrupolar strength $q_2$  plays  an important role in the 
above differences -- moreover when $q_1=0$  -- we shall examine closely
 the quadrupolar potentials. A significant quantity is  the  quadrupolar
  deviation,
\be 
(\psi_2-\psi^N_2)/\psi_2^N, \lb{dev}
\ee
that we shall study graphically.

In Fig. 1 we present the deviation  of $\psi_2=\psi^I_2$ on the
 plane $\vartheta=\pi/2$ for $a=1$, and different values of the constant
$\alpha$: 0.25 (top curve), 0.5, 0.75, 1.0 (bottom). We can have
 large deviations that asymptotically vanish. In Fig. 2 we show
 the deviation $\psi_2=\psi^{II}_2$ on the
 plane $\vartheta=\pi/2$ for  different values of the parameter
$a$: 0.25 (top curve), 0.5, 0.75, 1.0 (bottom). The case $a=1$ corresponds
to the ERQ solution.  It is not difficult to show using Letelier 
identity that in the 
limit $a=0,\; \psi^{II}=\psi^N$, in other words the ERQ solution
can continuously deformed in to $\psi^N$, fact that is confirmed in Fig. 2 for
the quadrupolar term.
 
\section{Solutions and differences.}

For our purposes, it is enough to consider the
 the multipolar series only up to the fourth order term, $\psi_4$.
 Since one of our  objectives  is the study
 of orbits of test particles moving around a deformed center 
of attraction, the inverse ``radial coordinate'',  
$u=R_S^{-1}=[m(1+x)]^{-1}$, will be useful. Moreover, all the metric 
functions will be computed in the 
plane $\vartheta=\pi/2$ up to the  sixth order in u; this order will
 be the adequated for consistence in the expansions.
 Also we shall disregard the dipolar moment, $q_1=0$ (we shall comeback to this point latter).
We find for the different functions
 $\hat\psi=\hat\psi_2+\hat\psi_3+\hat\psi_4$:
\ba
&&\hat\psi^N=-\frac{1}{2}q_2 u^3 -\frac{3}{2}mq_2 u^4-
\frac{3}{8}(10m^2q_2-q_4)u^5 \nn\\
&&\hspace{2cm}-\frac{5}{8}m(14m^2q_2-3q_4)u^6, \lb{psian}\\
&&\hat\psi^{ERQ}=-\frac{1}{2}q_2 u^3 -\frac{3}{2}mq_2 u^4-
\frac{3}{56}(64m^2q_2-7q_4)u^5 \nn\\
&&\hspace{2cm}-\frac{5}{56}m(80m^2q_2-21q_4)u^6,\lb{psiaerq}\\
&&\hat\psi^I=-\frac{1}{2}q_2 u^3 -\frac{3}{2}mq_2 u^4-
\frac{3}{8}[2(5-3a^2)m^2q_2+4(2\alpha-1)aq_3-q_4]u^5\nn\\
&&\hspace{2cm}-\frac{5}{8}m[2(7-9a^2)m^2q_2+60(2\alpha -1)amq_3-3q_4]u^6 ,\lb{psiai}\\
&&\hat\psi^{II}=-\frac{1}{2}q_2 u^3 -\frac{3}{2}mq_2 u^4-
\frac{3}{56}[(70-6a^2)m^2q_2-7q_4]u^5 \nn\\
&& \hspace{2cm}-\frac{5}{56}m[2(49-9a^2)m^2q_2-21q_4)u^6.\lb{pisaii}
\ea
The respective functions $\hat \gamma$ are:
\ba
&&\hat\gamma^N=-\frac{3}{4}u^4 -3m^2q_2u^5 
-\frac{1}{8}(3q^2_2 +70m^3q_2 -5mq_4)u^6,\lb{gaman}\\
&&\hat\gamma^{ERQ}=-\frac{3}{4}u^4 -3m^2q_2u^5 
-\frac{1}{56}(21q^2_2 +460m^3q_2 -35mq_4)u^6,\lb{gamaerq}\\
&&\hat\gamma^{I}=-\frac{3}{4}u^4 -3m^2q_2u^5 
-\frac{1}{8}(3q^2_2 +70m^3q_2-30a^2m^3q_2-5mq_4)u^6,\lb{gamai}\\
&&\hat\gamma^{II}=-\frac{3}{4}u^4 -3m^2q_2u^5 
-\frac{1}{56}(21q^2_2 +490m^3q_2-30a^2m^3q_2 \nn\\
&&\hspace{8cm}-35mq_4)u^6.\lb{gamaii}
\ea
The exact function $\gamma$ for the Schwarzschild  solution with Newtonian
multipoles can be found explicitly for the general case \cite{manko}, and in
the case of the multipoles $\hat\psi^{ERQ}$ in \cite{que}. These relations
 are rather
formidable and of not much use. An integral representation 
of the function  $\gamma$ for the ERQ solution can be found in \cite{letid}.
For the     solution $I$ we were able to find $\gamma$ in the general case.
For the solution $II$ up to the quadrupolar moment. These results are rather
 cumbersome an will be presented elsewhere. 

Note that to  eliminate conic singularities along the axis of symmetry we 
need to impose the boundary condition $lim_{r=0}\gamma=0$ outside the source.
The solutions with this boundary condition  are asymptotically flat.

Since our main goal is to study different 
solutions with the same Newtonian multipole moments we shall compute
the quantities,
\be
 \delta \psi = \psi -\psi^N,\;\; \delta \gamma = \gamma -\gamma^N.
 \lb{delta}
\ee
Note that $ \delta \psi= \delta\hat \psi$ and  $\delta \gamma= 
 \delta \hat \gamma$.
We find,
\ba
&& \delta \psi^{ERQ}=\frac{9}{28}m^2q_2u^5+\frac{45}{28}m^3q_2u^6,
 \lb{delperq}\\
&& \delta \psi^{I}=\frac{3}{4}a[3amq_2 -2(2\alpha-1)q_3]mu^5\nn\\
&&\hspace {3cm}+\frac{15}{4}a[3amq_2-2(2\alpha-1)q_3]m^2u^6, \lb{delpi}
\\
&& \delta \psi^{II}=a^2\delta \psi^{ERQ} . \lb{delpii}
\ea
And for the function $\gamma$,
\ba
&&\delta\gamma^{ERQ}=\frac{15}{28}m^3q_2u^6, \lb{delgerq}\\
&&\delta\gamma^{I}=\frac{5}{4}a[ 3amq_2 -2(2\alpha -1)q_3 ]m^2u^6, 
\lb{delgi}\\
&&\delta\gamma^{ERQ}=\frac{15}{28}a^2m^3q_2u^6. \lb{delgii}
\ea
These quantities will play and essential role in our analysis, we note that
the quadrupolar strength is the main parameter that appears in these
 differences. Also  in $\delta \psi^{I}$ and $ \delta\gamma^{I}$ we have 
the octopolar constant,
$q_3$. Note that the solution $\psi^I$ is highly asymmetric due to
 the weight constant $\alpha$. For particles orbiting a central
 body deformed by this highly asymmetric multipolar field  we might have 
 instabilities, for simplicity we shall take the symmetric case,
$\alpha=1/2$, in this case,
\ba
&&\delta \psi^{I}=\frac{9}{4}a^2mq_2u^5+\frac{45}{4}3a^3m^3q_2u^6, \lb{delpis}\\
&&\delta\gamma^{I}=\frac{15}{4}a^2m^3q_2u^6. \lb{delgis} 
\ea
In summary, we have that the discrepancies in the solutions appear
 in the fifth order
 of the inverse radial variable $u$ for the potential $\psi$ and the sixth
order for the potential $\gamma$,
\ba
&&\hat\psi=-\frac{q_2}{2}u^3-\frac{3}{2}mq_2u^4+
(-\beta m^2q_2+\frac{3}{8}q_4)u^5
+{\cal O}(u^6), \lb{psi5}\\
&&\hat\gamma=-\frac{3}{4}mq_2u^4-3m^2q_2u^5+{\cal O}(u^6). \lb{gam5}
\ea
with
\be\beta^N=\frac{15}{4}\  ,\  \beta^{ERQ}=\frac{24}{7}\  ,\  \beta^I=
\frac{3}{4}(5-3a^2)\  ,\ \beta^{II}=\frac{3}{56}(70-6a^2). \lb{beta1}
\ee
For the  differences up to the   fifth order in $u$ we have,
$ \delta \gamma=  \delta \hat \gamma =0 \lb{ dga0}$ and  
\be
\delta\psi =\delta\hat \psi = -\delta \beta m^2 q_2 u^5, \lb{dbeta1}
\ee
with
\be
\delta \beta^{ERQ}=-\frac{9}{28}, \;\;\delta\beta^{I}=- \frac{9}{4}a^2 ,
\;\; \delta \beta^{II}=-\frac{9}{28}a^2.  \lb{dbeta2}
\ee

For  general linear perturbations
 (not necessarily static axially symmetric)
of the Schwarschild black hole 
we have the Regge-Wheeler \cite{RW}  formalism. 
 One can  show \cite{vlapj} that the simple case of linear static axially
 symmetric perturbations in Weyl coordinates  can also be put in the
 Regge-Wheeler form, as long as, the  solutions do not have 
 conic singularities. 
In the framework of above mentioned  formalism one  can 
also show that the dipolar perturbations can be eliminated by a transformation
of coordinates as in the usual Newtonian case. This is not a trivial fact
since relativistic dipolar moments are invariant tensorial
quantities and as 
such cannot be made zero by a coordinate transformation. In the linear
approximation the dipolar perturbations turn to  be  gauge dependent.

\section{An evolution  equation for  differences.}

A first and simple approach to the problem of looking for the
 discrepancies in the orbits of test particles moving around a deformed
 attraction center  modeled with
 different relativistic gravitational fields (metrics)  with 
the same Newtonian limit,  is to compare  circular orbits by means of
 an  equation for the differences of the radial coordinate. 
The geodesic equation 
for the Weyl metric has three constants of motion, that in
 the plane $y=0$
determine completely the orbit of test particles:
\ba
&&e^{2\psi}\dot{t}^2-m^2e^{2\gamma-2\psi}(x^2-y^2)
(\frac{\dot{x}^2}{x^2-1}+     \frac{\dot{y}^2}{1-y^2})\nn\\
&&\hspace{2cm}-m^2e^{-2\psi}(x^2-1)(1-y^2)\dot{\varphi}^2=1, \lb{cm}\\
&&e^{2\psi}\dot{t}=l, \lb{ce}\\
&&e^{-2\psi}(x^2-1)(1-y^2)\dot{\varphi}=\frac{h}{m^2},\lb{cl}
\ea
where, as usual,  $\dot{t}=dt/ds$, etc., $l$ and $h$ are constants 
 related to the test particle energy and angular momentum, respectively.

From (\ref{cm})-(\ref{cl}) we find
 the equation of the orbits in the equatorial plane, $y=0$,
\be
l^2e^{-2\psi}-e^{2\gamma+2\psi}\frac{h^2}{m^2}\frac{x^2}{(x^2-1)^3}x'^2
-\frac{h^2}{m^2}\frac{e^{2\psi}}{x^2-1}=1, \lb{or1}
\ee
with $x'=\frac{dx}{d\varphi}$. It is more convenient to write this
equation in terms of the potentials,  $\hat\psi$ and
 $\hat\gamma$, and  the variable $u=R_S^{-1}=[m(1+x)]^{-1}$. Hence,
\be
u'^2+u^2e^{-2\hat\gamma}-\frac{l^2}{h^2}e^{-2(\hat\gamma+2\hat\psi)}+\frac{1}{h^2}
e^{-2(\hat\gamma+\hat\psi)}-\frac{2mu}{h^2}e^{-2(\hat\gamma+\hat\psi)}
-2mu^3e^{-2\hat\gamma}=0, \lb{or2}
\ee
that has the general form,
\be
u'^2=F(u,\hat{\gamma}(u),\hat{\psi}(u)), \lb{or3}
\ee
where
\be
F=e^{-2\hat\gamma}(-u^2+2mu^3+\frac{l^2}{h^2}e^{-4\hat\psi}
-\frac{1}{h^2}e^{-2\hat\psi}+\frac{2m}{h^2}ue^{-2\hat\psi}). \lb{F}
\ee
By differentiation  of (\ref{or3}) we obtain the general 
 equation for the difference of the inverse radial coordinate of
 equatorial orbits,
\be
(\delta u)'-\frac{1}{2u'}\frac{\partial F}{\partial u}\delta u=
\frac{1}{2u'}\left(
\frac{\partial F}{\partial \hat\gamma}\delta\hat\gamma
+\frac{\partial F}{\partial \hat\psi}\delta\hat\psi\right), \lb{diffeq}
\ee 
where $\delta\hat\gamma$ and $\delta\hat\psi$ represent
the differences defined in (\ref{delta}).

Note that
\be
\frac{\partial F}{\partial \hat\gamma}=-2F. \lb{Fga}
\ee

For circular orbits, $u=u_0$, ($R_S=1/u_0,$ the  constant radius),
we have $F=0$. From (\ref{Fga}) and  (\ref{diffeq}) we get
 \be
\delta u=-\left(\frac{\partial F}{\partial \hat\psi}/\frac{\partial
 F}{\partial u}\right)_{u_0}\delta\hat\psi. \lb{delu01} 
\ee
 First, we shall compute the  leading term  in $u$ of this last equation, from 
(\ref{dbeta1}) we find,
\be
 \delta u=\frac{1}{3}(1-2l^2) q_2 u^4_{0}\delta \beta . \lb{delu02}
\ee

This last equation in terms of the Schwarzschild like  radial coordinate $R_S$ reads,
\be
-\delta R_S =\frac{1}{3}(1-2l^2)q_2\frac{\delta \beta}{R_{S}^2}. \lb{delu02n}
\ee
Then the  difference decays with the square  of the radius. It is also instructive to express  equation  (\ref{delu02}) in terms of  
 the dimensionless parameter,
\be
\epsilon=\frac{\mathrm {half\  Schwarzschild\  radius}}
{\mathrm {orbit\  radius}}. \lb{eps}
\ee 
We get,
\be
\delta u=\delta\beta(1-2l^2)\frac{q_2}{ 3 m^3}u_0  \epsilon^3. \lb{delu03}
\ee
For planets like Earth, Mercury, and Mars  $\epsilon$ is a
 little parameter  $\epsilon \sim 10^{-8}$. But, for
a small test body orbiting around a neutron star we can have
  $\epsilon \sim 10^{-1}$. In this last case the multipolar fields can be
 originated, for instance,  by polar  jets of matter ejected  by  the neutron
 star.  
Finally, we want to comment that  Eq. (\ref{delu01}) 
gives a simple expression involving the exact metric.

\section{ The perihelion shift.}

To  compare the different multipole expansions we  shall  study
 the perihelion shift  of a test particle orbiting in the
 equatorial plane of the deformed ``black hole''.                                
 In  the order of approximation used in the present work 
we can put  $e^{\hat\psi}=1+\hat \psi $ and
$e^{\hat\gamma}=1+\hat \gamma$ in (\ref{or2}) with no error, 
\ba
&&u'^2+u^2-\frac{l^2-1}{h^2}-\frac{2mu}{h^2}-2mu^3= \nn\\
&&2\hat\gamma\left(u^2-\frac{l^2-1}{h^2}-\frac{2mu}{h^2}-2mu^3\right)
+2\hat\psi\left(-\frac{2l^2}{h^2}+\frac{1}{h^2}-\frac{2mu}{h^2}\right).
\lb{ora1}
\ea
 Doing  $\hat\psi=\hat\gamma=0$ in the previous equation 
 we obtain the well known equation for
 the motion of a test  particle in the  Schwarzschild spacetime.

 Since the first term in the series expansion of the functions
   $\hat\gamma$  and  $\hat\psi$  are proportional to $u^4$
 and  $u^3$, respectively  [cf. Eqs.
 (\ref{psian})--(\ref{gamaii})], to use the orbital equation
 (\ref{ora1}) without error we must limit the
expansion of $\hat\gamma$ and $\hat\psi$  to the order seven and  five in the variable $u$, respectively.
Therefore we shall  consider all  the series up to the  fifth order
 in $u$, that is the order wherein  the first discrepancies
 between the different multipole expansions appear. 
From Eq.  (\ref{ora1}) we
obtain,
\be
u'^2+u^2=[l^2-1 +2mu+2mh^2u^3
+\frac{2}{3}qm^3u^3+\frac{1}{2}rm^4u^4 +\frac{2}{5}sm^5u^5]/h^2, \lb{ora2} 
\ee
where,
\ba
&&q=\frac{3}{2}\left(2(l^2-1)+1\right)\frac{q_2}{m^3}, \lb{defq}\\
&&r=\left(15(l^2-1)+10\right)\frac{q_2}{m^3},\lb{defr}\\
&&s=\frac{5}{2}\left[\left((4\beta+6)(l^2-1)+2\beta+9\right)\frac{q_2}{m^3}
-\frac{3}{4}(2l^2-1)\frac{q_4}{m^5}\right]. \lb{defs}
\ea
Note that $q$, $r$, and  $s$ are dimensionless quantities.
The semilatus rectum for the  usual  Keplerian orbit of a particle 
attracted by a central body with no multipolar deformations is,
$p=h^2/m.$ 
We shall use
 a dimensionless parameter similar to the homonymous one of the 
preceding section,
\be
\epsilon=\frac{m^2}{h^2}=\frac{m}{p}=
\frac{\mathrm {half\  Schwarzschild\  radius}}{\mathrm 
{semilatus\  rectum}}.
\lb{defep2} \ee 
Note that for a circular orbit with $u=u_0$ we have $u_0=1/p$.
It is  enlightening to write the equation (\ref{ora2}) in terms of the new
dimensionless variable $w=pu$,
\be
w'^2+w^2=p^2\frac{l^2-1}{h^2}+2w+2\epsilon w^3+\frac{2}{3} q\epsilon^2 w^3
+\frac{1}{2}r\epsilon^3 w^4+\frac{2}{5}s\epsilon^4 w^5. \lb{ora3}
\ee
By derivation we obtain,
\be
w''+w=1+3\epsilon w^2+q\epsilon^2 w^2+r\epsilon^3 
w^3+s\epsilon^4 w^4. \lb{ora4}
\ee
We remind that the  discrepancies between the different multipole expansions
are controlled  by the  parameter $\beta$ that it only 
appears in $s$ (last term).
We shall  search a solution of equation (\ref{ora4}) in the form of a 
series expansion up to the fourth order in $\epsilon$,
\be
w=w_0+ w_1\epsilon+w_2\epsilon^2 +w_3 \epsilon^3 +w_4\epsilon^4, \lb{sw}
\ee
From (\ref{ora4}) we find  the system of linear equations,
\ba
&&w''_0+w_0=1, \nn\\
&&w''_1+w_1=3w_0^2, \nn\\
&&w''_2+w_2=6w_0w_1+qw_0^2, \nn \\
&&w''_3+w_3=3(2w_0w_2+w_1^2)+2qw_0w_1+rw_0^3, \nn\\
&&w''_4+w_4=6(w_0w_3+w_1w_2)+q(2w_0w_2+w_1^2)+3rw_0^2w_1+sw_0^4. \lb{weqs}
\ea
The solution of this system can be expressed in 
terms of elementary functions,
we find,
\ba
&&w_0=1+{\rm e} \cos\varphi, \nn\\
&&w_1=3(1+{\rm e} ^2/2)+3{\rm e} \varphi\sin\varphi
-({\rm e} ^2/2)\cos{2\varphi}, \nn\\
&&w_2=9(2+{\rm e} ^2)+q(1+{\rm e}^2/2)-(9/2){\rm e} \varphi^2\cos\varphi
-(3{\rm e} ^2+q{\rm e} ^2/6)\cos{2\varphi} \nn\\
&&\hspace{1cm}+(3{\rm e}^3/16)\cos{3\varphi}+(27{\rm e}/2+15{\rm e} ^3/4+q{\rm e} )\varphi\sin\varphi-3{\rm e} ^2\varphi\sin{2\varphi},\nn\\
&&w_3=135 + 81{\rm e}^2 + 57{\rm e}^4/8 + 12q + 6{\rm e}^2q +
 r + 3{\rm e}^2r/2 - (81{\rm e}\varphi^2/2\nn\\
&& \hspace{1cm} + 45{\rm e}^3\varphi^2/4 + 3{\rm e}q\varphi^2)\cos{\varphi}
 -  (27{\rm e}^2 +59{\rm e}^4/16 + 2{\rm e}^2q + {\rm e}^2r/2 \nn\\
&&\hspace{1cm}-9{\rm e}^2\varphi^2)\cos{2\varphi} + 
  (9{\rm e}^3/4 + {\rm e}^3q/8 - {\rm e}^3r/32)\cos{3\varphi} -
 \frac{1}{16}{\rm e}^4\cos{4\varphi} + \nn\\
&&\hspace{1cm}  (189{\rm e}\varphi/2 + 135{\rm e}^3\varphi/4 
+ 9{\rm e}q\varphi+
 5{\rm e}^3q\varphi/2 + 3{\rm e}r\varphi/2 + 
     3{\rm e}^3r\varphi/8  \nn\\
&& \hspace{1cm}- 9{\rm e}\varphi^3/2)\sin{\varphi}- (63{\rm e}^2\varphi/2
 + 15{\rm e}^4\varphi/4 + 2{\rm e}^2q\varphi)\sin{2\varphi}\nn\\
&&\hspace{1cm}  + \frac{27}{16}{\rm e}^3\varphi \sin{3\varphi},\nn \\
&&w_4=1134 + 810{\rm e}^2 + 513{\rm e}^4/4 + 135q + 81{\rm e}^2q
 + 57{\rm e}^4q/8 + 2q^2 +   {\rm e}^2q^2 \nn\\
&& \hspace{1cm} + 15r + 18{\rm e}^2r + 15{\rm e}^4r/8 + s +
 3{\rm e}^2s+  3{\rm e}^4s/8  
 - (2997{\rm e}\varphi^2/8 \nn\\
&&\hspace{1cm} + 1215{\rm e}^3\varphi^2/8 + 225{\rm e}^5\varphi^2/32 + 81{\rm e}q\varphi^2/2 +  45{\rm e}^3q\varphi^2/4 + {\rm e}q^2\varphi^2/2 \nn\\
&&\hspace{1cm}+ 9{\rm e}r\varphi^2/2 +9{\rm e}^3r\varphi^2/8 - 
     27{\rm e}\varphi^4/8)\cos\varphi + (-270{\rm e}^2 - 531{\rm e}^4/8 \nn\\
&&\hspace{1cm}- 27{\rm e}^2q - 59{\rm e}^4q/16 - {\rm e}^2q^2/3
 - 6{\rm e}^2r - 31{\rm e}^4r/32 - {\rm e}^2s - {\rm e}^4s/6 + \nn\\
&&\hspace{1cm}135{\rm e}^2\varphi^2 + 45{\rm e}^4\varphi^2/2 + 9{\rm e}^2q\varphi^2)\cos{2\varphi} + 
  (27{\rm e}^3 + 711{\rm e}^5/256 + 9{\rm e}^3q/4 +\nn\\
&&\hspace{1cm} {\rm e}^3q^2/48 + 3{\rm e}^3r/16 - 
     {\rm e}^3s/8 - 243{\rm e}^3\varphi^2/32)\cos{3\varphi} - 
  (9{\rm e}^4/8 + {\rm e}^4q/16  \nn\\
&&\hspace{1cm} - {\rm e}^4r/32 + {\rm e}^4s/120)\cos{4\varphi} + 
  \frac{5}{256} {\rm e}^5\cos{5\varphi} + (6237{\rm e}\varphi/8 +5{\rm e}^3 q^2
\varphi/12\nn\\
&&\hspace{1cm} + 2835{\rm e}^3\varphi/8 + 1455{\rm e}^5\varphi/64 + 
     189{\rm e}q\varphi/2 + 135{\rm e}^3q\varphi/4 + 3{\rm e}q^2\varphi/2\nn\\
&&\hspace{1cm}+  33{\rm e}r\varphi/2 + 69{\rm e}^3r\varphi/8 +
 2{\rm e}s\varphi +3{\rm e}^3s\varphi/2 - 243{\rm e}\varphi^3/4 - 135{\rm e}^3\varphi^3/8\nn\\
&&\hspace{1cm} - 9{\rm e}q\varphi^3/2)\sin{\varphi}
 -(675{\rm e}^2\varphi/2 + 627{\rm e}^4\varphi/8 +63{\rm e}^2q\varphi/2+ 15{\rm e}^4q\varphi/4 \nn\\
&&\hspace{1cm}  +{\rm e}^2q^2\varphi/3 + 9{\rm e}^2r\varphi/2 + 3{\rm e}^4r\varphi/8 -18{\rm e}^2\varphi^3)\sin{2\varphi}+ (891{\rm e}^3\varphi/32   \nn\\
&&\hspace{1cm} + 135{\rm e}^5\varphi/64 + 27{\rm e}^3q\varphi/16 - 9{\rm e}^3r\varphi/32)
   \sin{3\varphi}- \frac{3}{4}{\rm e}^4\varphi \sin{4\varphi}.
\lb{wss}
\ea
  Only the function $w_4$ contains 
the parameter $s$.

The perihelion, as well as,  the aphelion of the test particle orbit are 
 given by  extremals  of the function $u$ or $w$. 
In (\ref{wss}) we  chose the constants of integration to have
$\varphi=0$ in the position of the perihelion at any  order in 
$\epsilon$. The  aphelion is nearby $\varphi=\pi$ and the next 
perihelion closed to $\varphi=2\pi$. The integration constant $\rm e$ is the 
orbit eccentricity.

We shall begin the computation of the perihelion shift in the lower order  
in $\epsilon$. We have
\be
w=w_0+\epsilon w_1,
\ee
where $w_0$ and $w_1$ are given in (\ref{wss}). We verify that, $w'(0)=0$.
Therefore,   the  perihelion is located the in the angular
 position   $\varphi=0$; the next perihelion is in
$\varphi=2\pi+L$, where $L$ is a small unknown quantity. We have by
definition,  $w'(2\pi+L)=0$. Expanding  the function
 $w'(\varphi)$ up to the first order in  $L$ in the neighborhood
 of $2\pi$, we find
\be
w'(2\pi+L)-w'(2\pi)=Lw''(2\pi)+{\cal O}(L^2). \lb{per0} 
\ee
We obtain the advance of the perihelion 
up to the first order in $\epsilon$,
$\;\; L=6\pi\epsilon$, i.e.,  the usual result for a particle orbiting
 in the Schwarzschild metric.
Now, $w$ up to  the  four order  in $\epsilon$  is
 given by the series (\ref{sw}). 
From the previous result we know that the series expansion of  $L$ 
begins  with the  order $\epsilon$.
Thus,  we will look for  a constant  $L$  of the form, 
\be
L=L_1\epsilon+L_2\epsilon^2+L_3\epsilon^3+L_4\epsilon^4, \lb{L}
\ee
solution of the equation
\be
w'(2\pi+L)-w'(2\pi)=Lw''(2\pi)+\frac{L^2}{2}w'''(2\pi)
+\frac{L^3}{6}w''''(2\pi)+{\cal O}(L^4). \lb{per1}
\ee 
Since the fifth derivative is of order $\epsilon$ we
 cut the expansion in the  fourth derivative . Also, we
note that $w'(0)=w'(2\pi+L)=0$.
By identification of the coefficients of the different 
powers of $\epsilon$ in the series expansion of the
two member of Eq. (\ref{per1})     we obtain,
\ba
&& \hspace{-1cm} L_1=6\pi, \nn \\
&&\hspace{-1cm} L_2=\frac{1}{2}\pi(90+15{\rm e} ^2+4q),\nn\\
&&\hspace{-1cm} L_3=\pi(1620+450{\rm e}^2 +120q +20{\rm e}^2q+12r
+3{\rm e}^2 r)/4,\\
&&\hspace{-1cm} L_4=\pi(379080+145800{\rm e}^2+ 7065{\rm e}^4+ 38880q
+ 10800{\rm e}^2q+480q^2 \nn\\
&&\hspace{1cm}+80{\rm e}^2q^2+4896r+2088{\rm e}^2r+384s
 +288{\rm e}^2s)/96.\lb{Lr}
\ea
We see that the discrepancies between the different multipole expansions 
are reflected in $L_4$ via  $s$. So until the order 
$\epsilon^3$  we have the same perihelion shift for the
multipole expansions considered in the present paper.
Moreover, Eqs. (\ref{L}) and (\ref{Lr}) with $q=r=s=0$ gives us the
 perihelion shift
 of an orbit in Schwarzschild geometry up to the fourth order in $\epsilon$.
 We have that  the contribution to the perihelion due to deformation of the
spherically symmetric 
  attraction center appears in  the second order in $\epsilon$; a fact
 that  is well known in Newtonian theory.
  
Now, let us return to the difference equation (\ref{diffeq}).
For $\hat\psi$ and $\hat\gamma$ we use the solutions up to the five order
 in $u$. Therefore we have, $\delta\hat\gamma=0$ 
and $\delta{\psi}$ given by (\ref{dbeta1}).
The equation (\ref{diffeq}) reduces to,
\be
u'(\delta u)'-\frac{1}{2}\left(\frac{2}{p}-2u+6mu^2+{\cal O}(\epsilon)\right)\delta u
=\delta\beta(2l^2-1)\frac{q_2}{m^3}p^3\epsilon^4u^5. \lb{diffeqor1}
\ee
 In this equation we can replace the solution $u$, or equivalently $w$,
 by his first order approximation $w_0=1+{\rm e} \cos\varphi$. We find for the lower order in $\epsilon$,
\be
{\rm e} \sin \varphi (\delta u)'-\left[{\rm e} \cos\varphi -\frac{3m}{p}(1+{\rm e} \cos\varphi)^2\right]\delta u=-\delta\beta(2l^2-1)
\frac{q_2}{m^3}\frac{1}{p}\epsilon^4(1+{\rm e} \cos\varphi)^5. \lb{diffeqor2}
\ee
By letting $ \rm{e} \rightarrow 0 $ in the previous equation we recover the equation for the differences of circular orbits, Eq. (\ref{delu03}).
Since  Eq. 
(\ref{diffeqor2})  is a 
 linear  first order differential equation, its  solution 
 has a simple   integral representation that in this case is rather 
 useless due to the fact that the integrals are not elementary.
The nonhomogeneous part of the solution is proportional to $\epsilon^4$. Fact
that confirms that the  
first discrepancies between the different expansions are proportional to $\epsilon^4$.

\section{Discussion}

From  the study of particular cases,  we believe,
 that are representative of
different possible series expansions of the metric functions for
  Weyl solutions that have the same Newtonian limit. We found  a strong
 indication that
the use of the right relativistic multipolar expansion to describe a
deformed body need to be considered only in the case of a 
strong gravitational
regime like bodies orbiting close (a few Schwarzschild radius)  around
a very  compact object like a neutron star.  
For usual planetary motion, in particular for the perihelion shift,
 the effect of having different  multipolar expansions 
with the same Newtonian limit can be completely ignored. 
Also when studying the stability of orbits (chaos) of particles moving
 around deformed bodies, one of us \cite{guelet} found no difference in 
the trajectories for Newtonian and ERQ deformations.
Recently, the same effect was study on the motion of a gyroscope
 \cite{herrera}. 

\vspace{1cm}

{\bf \noindent
Acknowledgments}

P.S.L. thanks  CNPq and FAPESP for financial support and  S.R.  Oliveira
 for discussions, also acknowledges the warm hospitality of 
the ``Laboratoire''.

\newpage
\begin{center}
\begin{figure}
\epsfig{width=3in,height=5in,angle=-90,file=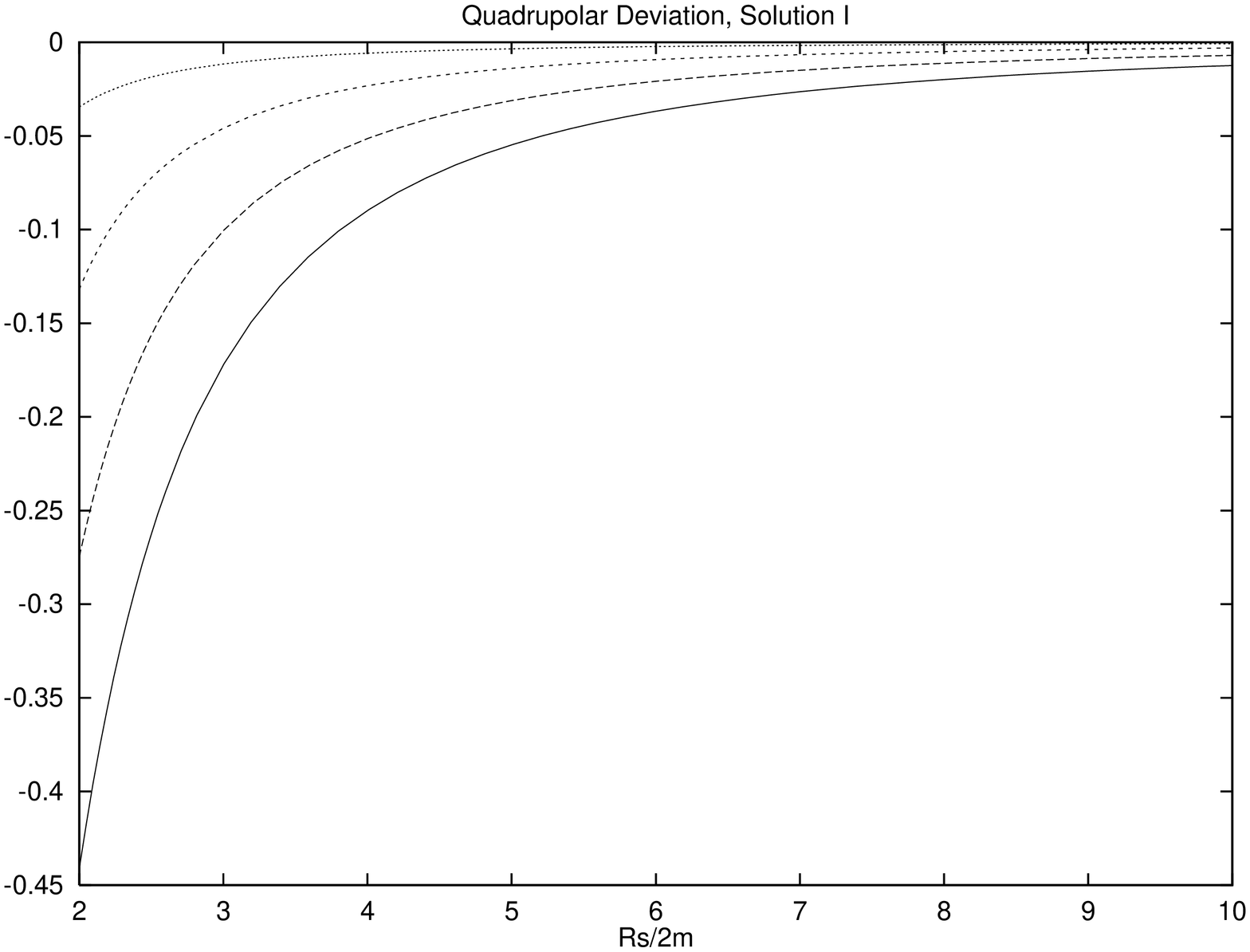}
 \caption{We present the deviation for $\psi_2=\psi^I_2$ on the
 plane $\vartheta=\pi/2$ for $a=1$, and different values of the constant
$\alpha$: 0.25 (top curve), 0.5, 0.75, 1.0 (bottom).}
\end{figure}
\end{center}

\newpage
\begin{center}
\begin{figure}
\epsfig{width=3in,height=5in,angle=-90,file=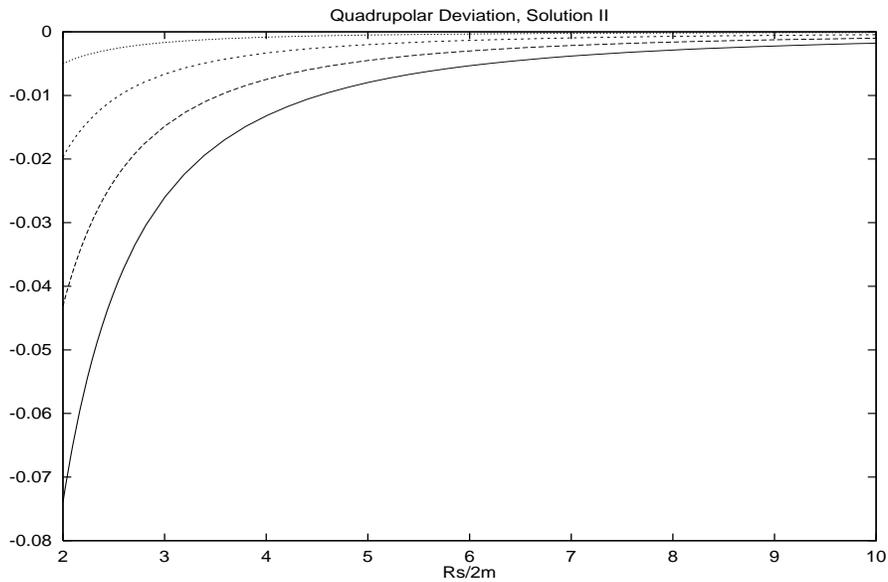}
 \caption{We show the deviation of $\psi_2=\psi^{II}_2$ on the
 plane $\vartheta=\pi/2$ for  different values of the parameter
$a$: 0.25 (top curve), 0.5, 0.75, 1.0 (bottom).
 The case $a=1$ corresponds to the ERQ solution.  }
\end{figure}
\end{center}

\end{document}